\documentclass[aps,epsfig]{revtex4}
\usepackage{graphicx}
\begin{document}
\def\be{\begin{equation}}
\def\ee{\end{equation}}
\def\bfi{\begin{figure}}
\def\efi{\end{figure}}
\def\bea{\begin{eqnarray}}
\def\eea{\end{eqnarray}}

\title{Crossover between Ising and XY-like behavior in the off-equilibrium
kinetics of the one-dimensional clock model}

\author{Natascia Andrenacci$^\S$, Federico Corberi$^\dag$, and Eugenio Lippiello$^\ddag$}
\affiliation {Dipartimento di Fisica ``E.Caianiello'' and CNR-INFM
Istituto Nazionale di Fisica della Materia,
Universit\`a di Salerno,
84081 Baronissi (Salerno), Italy}

\begin{abstract}
We study the phase-ordering kinetics following a quench to a final temperature $T_f$ 
of the one-dimensional $p$-state clock model.
We show the existence of a critical value $p_c=4$, where the properties of
the dynamics change. At $T_f=0$, for $p\le p_c$ the dynamics
is analogous to that of the kinetic Ising model, characterized by 
Brownian motion and annihilation of 
interfaces. Dynamical scaling is obeyed with
the same dynamical exponents and scaling functions of the Ising model.
For $p>p_c$, instead, the dynamics is dominated
by a texture mechanism analogous to the one-dimensional XY model, 
and dynamical scaling is violated. During the phase-ordering process 
at $T_f>0$, before equilibration occurs, a cross-over
between an early XY-like regime and a late Ising-like 
dynamics is observed for $p>p_c$. 
\end{abstract}

\maketitle

\S andrenacci@sa.infn.it

\dag corberi@na.infn.it \ddag lippiello@sa.infn.it 


\section{Introduction} \label{intro}

After quenching a ferromagnetic system to a low temperature phase, 
relaxation towards the 
new equilibrium state is realized by a progressive phase-ordering~\cite{Bray94}.
The specific mechanisms involved in the coarsening phenomenon
depend on the presence and on the nature of topological defects 
seeded by the disordered initial configuration which,
in turn, are determined by the space dimensionality $d$ and
the number of components $N$ of the order parameter.
For $N<d$ defects
are spatially extended; in this case coarsening is driven by 
reducing the typical curvature of the defect core, removal of
sharp features and shrinking of domain bubbles or vortex loops.
Systems with $N=d$ are characterized by
the presence of stable localized topological defects and 
the ordering process occurs by
mutual defect-antidefect annihilation.
This is the case of the Ising chain quenched to
a final temperature $T_f=0$, where 
up and down domains are separated by point-like
interfaces performing Brownian walks.
When $N=d+1$, such as in the one dimensional XY model,
the kinetics is characterized by textures, 
spatially extended defects without a core,
along which the order parameter rotates by $2\pi$.
Growth of the typical size of textures
is a relevant mechanism at work in these systems.
Finally, for $N>d+1$ topological defects are unstable and
the dynamics is solely driven by the reduction 
of the excess energy related to the smooth rotations 
of the order parameter. 

In any case, the development of order is associated to the growth
of one or more characteristic lengths, with laws that, 
besides the specific mechanisms discussed
above, depend on the conservation laws of the dynamics.

Generally, the late stage is characterized by
dynamical scaling. This implies that 
a single characteristic length
$L(t)$ can be associated to the development of order 
in such a way that configurations of
the system are statistically independent of time when
lengths are measured in units of $L(t)$.  
The characteristic length usually has a power law growth 
$L(t)\propto t^{1/z}$. 
In systems with a non-conserved order parameter one
generally finds $z=2$. In particular, this value
is provided by the exact
solution of the kinetic Ising chain~\cite{Glauber63}
quenched to zero temperature.

However, there are cases where dynamical scaling is
violated, notably the XY model in $d=1,2$. In $d=1$
this is related~\cite{Rutenberg95} to the presence of two lengths
$L_w(t)$ and $L_c(t)$,
associated to the texture length and to the
texture-antitexture distance,
growing with different exponents $z=4$ and $z=2$ respectively.

In this Article, we investigate the interplay between
two coarsening mechanisms, point-like defect annihilation
and texture growth, in the phase-ordering
kinetics of the one dimensional $p$-state clock model.
This spin system reduces to the Ising model for $p=2$
and to the XY model for $p=\infty $.
We study how the model with generic $p$ interpolates
between these limiting cases which,
as discussed above, behave
in a radically different way.
In doing that, we uncover the existence of a critical
value $p_c=4$, where the properties of
the dynamics change abruptly. For $p\le p_c$ the dynamics
at $T_f=0$ is characterized by Brownian motion 
and annihilation of 
interfaces between domains,
as in the Ising model. One has dynamical scaling with
the same dynamical exponents and, interestingly,
the same scaling functions of the Ising model.
For $p>p_c$, instead, the dynamics is dominated
by a texture mechanism analogous to the case
with $p=\infty $, and dynamical scaling is violated. 

In $d=1$ there is
no possibility of ergodicity breaking except at $T=0$.
At any finite temperature the equilibrium state is disordered
with a vanishing magnetization and a coherence length $\xi (T)$ that
diverges in the $T\to 0$ limit.
If the system is quenched to a 
sufficiently low temperature 
one has a coarsening phenomenon in a pre-asymptotic transient 
until the growing length associated to the development of order
becomes comparable with $\xi (T_f)$. Since $\xi (T_f)$ diverges as
$T_f\to 0$ the phase-ordering stage can be rather long.
In this regime we show that activated processes
restore, after a characteristic time $\tau _p^{cross}(T_f)$,
the Ising behavior also in the cases with $p>p_c$. 

This paper is organized as follows:
In Sec.~\ref{model} we introduce the model and define the
observable quantities that will be considered.
In Sec.~\ref{num} we present the outcome of numerical
simulations of the model with different $p$.
In particular, quenches to $T_f=0$ or
to $T_f>0$ will be discussed in Secs.~\ref{Tzer} and \ref{Tnzer}, respectively.
Here we enlighten the crossover between the Ising and the XY
universality class and provide an argument 
explaining its microscopic origin. 
A summary and the conclusions are contained in Sec.~\ref{concl}.

\section{Model and observables} \label{model}

The $p$-state clock model in one dimension is defined by the Hamiltonian
\be
H[\sigma ]=-J\sum _{i=1}^{\cal N}\vec \sigma _i \cdot \vec \sigma _{i+1}=
-J\sum _{i=1}^{\cal N}cos(\theta _i-\theta _{i+1}),
\label{hamiltonian}
\ee
where $\vec \sigma _i$ is a two-components unit vector spin pointing along  
one of the directions
\be
\theta _i=\frac{2\pi}{p} n_i, 
\label{theta}
\ee
with $n_i \in \{1,2,...,p\}$, $i=1,...,{\cal N}$
are the sites of the lattice and we assume periodic 
boundary conditions $\theta _{{\cal N}+1}=\theta _1$.
This spin system is 
equivalent to the Ising model for $p=2$ and to the XY model 
for $p\to \infty $. 
In $d=1$ the system is ergodic except at $T=0$.
At any finite temperature the equilibrium state is disordered
with a vanishing magnetization and a coherence length $\xi (T)$ that
diverges in the $T\to 0$ limit.

We consider a system initially prepared in an
high temperature uncorrelated state and then quenched, at time $t=0$,
to a lower final temperature $T_f$. 
The dynamics is characterized by the ordering of the system over
a characteristic length growing in time until, at time $\tau _p^{eq}(T_f)$ it becomes
comparable to $\xi (T_f)$. At this point the final equilibrium state
at $T_f$ is entered. Quenching to $T_f=0$, since $\xi (0)=\infty $,
one has $\tau _p^{eq}(T_f)=\infty$; therefore
an infinite system never reaches equilibrium and the phase-ordering kinetics
continues indefinitely. 
If the system is quenched to a 
sufficiently low temperature, since $\xi (T_f)$ is very large, 
the same behavior, as for $T_f=0$, can be observed over the time window 
$t<\tau _p^{eq}(T_f)$. 
 
The power law growth of the characteristic size of ordered regions 
depends on the specific mechanisms at work in the kinetic
process. In the 1d Ising model with non conserved order parameter, i.e.
single spin flip dynamics, ordering is determined by the Brownian motion
of the interfaces between up and down domains, which annihilate upon meeting.
This leads to 
\be
L(t)\sim t^{\frac{1}{z}},
\label{ldit}
\ee
with $z=2$. The same value is also expected~\cite{Leyvraz86} for $p\le 4$.

The situation is different in the
XY model in $d=1$. Here the order parameter is a vector which can
gradually rotate with a low energy cost. A  
smooth $2\pi $ rotation 
of the phase $\theta $ is called a texture when the rotation is clockwise,
or antitexture when it is counterclockwise. The length over which this phase winding 
occurs will be denoted by $L_w(t)$. After a quench from a disordered
state textures and antitextures are formed with equal probability. 
Then, there are points
where the rotation of $\theta $ changes direction
and the phase decohere. We denote with
$L_c(t)$ the characteristic length over which the phase remains coherent.
It was shown~\cite{Rutenberg95} that $L_w(t)$
and $L_c(t)$ grow with a power law~(\ref{ldit})    
but with different exponents. Specifically one has $z=4$ for
$L_w(t)$ and $z=2$ for $L_c(t)$. The existence of these two
lengths is at the heart of the scaling violations of the XY model.

Characteristic lengths can be estimated from the
knowledge of the two-points equal time correlation function
\be
G(r,t)=\langle \vec \sigma_i(t) \cdot \vec \sigma_{i+r}(t) \rangle,
\label{gdir}
\ee
where $\langle \dots \rangle$
means an ensemble average, namely taken over different initial conditions
and thermal histories.
Due to space homogeneity, $G(r,t)$ does
not depend on $i$.
If there is a single characteristic length in the system, 
one has dynamical scaling~\cite{Bray94}, which implies
\be
G(r,t)=g(x),
\label{scalgferro}
\ee  
where $x=r/L(t)$.
In the Ising model one finds~\cite{Glauber63}
\be
g(x)=erfc \{x\}.
\label{struttising}
\ee
with $L(t)=\sqrt {2t}$.
For small $x$ one has the Porod linear behavior $1-g(x)\sim x$, which is 
expected for systems with sharp interfaces~\cite{Bray94}.
From Eq.~(\ref{scalgferro}) one can extract a quantity $L_G(t)$ proportional to $L(t)$
from the condition
\be
G\left (L_G(t),t\right )=\frac{1}{2},
\label{halfheight}
\ee
namely as the half-height width of $G(r,t)$. 
In the XY model, 
$G(r,t)$ still obeys Eq.~(\ref{scalgferro}), with $x=r/L_w(t)$,
$L_w(t)=2^{3/4}(\pi t)^{1/4}$,
and~\cite{Bray94}
\be
g(x)=\exp \left \{-\frac{x^2}{\xi _i}\right \}.
\label{xystrutt}
\ee
Here $\xi _i$ is the correlation length of the initial condition which,
for a quench from a disordered state, is of the order of the lattice spacing.
The Porod law is not obeyed, since instead of sharp interfaces
one has smooth textures. 
Note that $G(r,t)$ has a scaling form, although dynamical 
scaling is violated. Scaling violations can be evidenced 
by considering different quantities as, for instance, the autocorrelation
function
\be
C(t,s)=\langle \vec \sigma_i (t)\cdot \vec \sigma_i (s) \rangle.
\label{autocorr}
\ee
In the Ising model this quantity can be cast in scaling form~\cite{Glauber63}
\be
C(t,s)=h(y),
\label{ccs}
\ee
where $y=t/s$ and
\be
h(y)=\frac{2}{\pi}\arcsin \sqrt {\frac{2}{1+y}}.
\label{ccsc}
\ee
In the XY model, instead,
one finds~\cite{Rutenberg95} the stretched exponential behavior
\be
C(t,s)=\exp \left \{-\frac{1}{\xi _i \sqrt \pi}s^{\frac{1}{2}}
\left [ 2(y+1)^{\frac{1}{2}}-(2y)^{\frac{1}{2}}-\sqrt 2 \right ] \right \},
\label{cxy}
\ee
This expression cannot be cast in a scaling form, as for the Ising model,
revealing the absence of dynamical scaling.

\section{Numerical results}\label{num}

In the following we will present the numerical results. 
Setting $J=1$, for each case considered we simulated a 
string of $10^4$ spins with periodic boundary conditions
and different values of $p$ ranging from $p=2$, corresponding
to the Ising model, to $p=\infty$, corresponding to the XY model. 
We consider a single spin flip dynamics regulated by
transition rates
\be
w\{[\sigma]\to [\sigma']\}=w_p(\frac {\Delta E}{T})=
\frac{2}{p}\,\,\,\frac{\exp \left (\frac {-\Delta E}{2T}\right )}
{\exp \left (\frac {\Delta E}{2T}\right )+\exp \left (\frac {-\Delta E}{2T}\right )}.
\label{metropolis} 
\ee
Here $[\sigma]$ and $[\sigma']$ are the spin configurations before and
after the move, differing at most by the value of the spin on a randomly chosen
site, $\Delta E=H[\sigma']-H[\sigma]$,
and we have set the Boltzmann constant to unity.
The transition rates~(\ref{metropolis}) are a generalization of Glauber
transition rates to the $p$-state spins of the clock model. 
They reduce to the usual Glauber
transition rates $w\{[\sigma]\to [\sigma']\}=(1/2)[1+\tanh (-\Delta E/2T)]$
for $p=2$. The factor $2/p$ in Eq.~(\ref{metropolis}) ensures 
that all spin values have the same probability $1/p$ when $\Delta E=0$.

An average over $10^4$ realizations is made for each simulation. 
The statistical errors in the data reported in the figures are always
smaller than the dimension of the symbols or the thickness of the lines.

\subsection{Quenches to $T_f=0$.} \label{Tzer}

\begin{figure}
\vspace{2cm}
    \centering
   \rotatebox{0}{\resizebox{.5\textwidth}{!}{\includegraphics{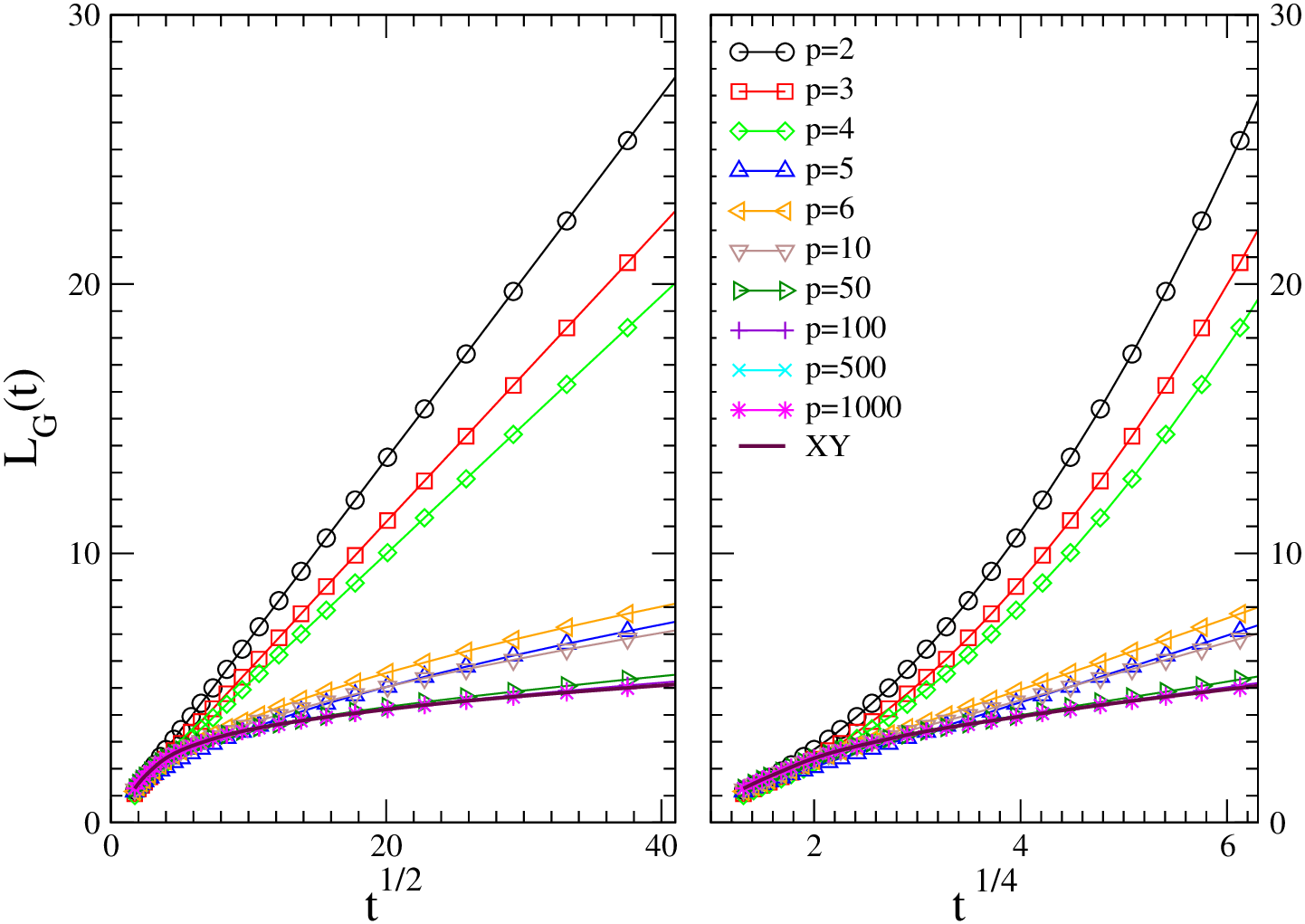}}}
    \caption{(Color online) 
     The characteristic length $L_G(t)$ is plotted against $t^{1/2}$ (left panel)
             and $t^{1/4}$ (right panel).} 
\label{figlengthT0}
\vspace{2cm}
\end{figure}

Let us start with quenches to $T_f=0$, by illustrating the behavior of  
the characteristic length $L_G(t)$ defined in Eq.~(\ref{halfheight}). 
In Fig.~\ref{figlengthT0} this quantity
is plotted against $t^{1/2}$ (left panel) or against $t^{1/4}$ (right panel),
for several values of $p$ ranging from $p=2$ to $p=\infty $. 

This figure
shows that $L_G(t)$ has an asymptotic power law growth, as in Eq.~(\ref{ldit}), 
for every value of $p$.
However,  the dynamic exponent $z$ radically changes going from 
$p\le p_c$, where one has values very well consistent with $z=2$
(best fits yield $1/z=0.49\pm 0.01$ for $p=2,3,4$), to $p>p_c$
where $z=4$ is found with good accuracy 
(we find $1/z=0.27\pm 0.01, 0.27\pm 0.01, 0.25\pm 0.01$ for $p=5,6,10$). 
We recall that these are the values
found in the Ising model and in the XY model. The behavior of $L_G(t)$,
then, indicates a crossover from Ising to XY behavior 
upon crossing $p_c=4$. We will see in the following that this
is confirmed by the analysis of other dynamical quantities.
Before doing that, however, let us discuss which is the microscopic
mechanism at the basis of this crossover.

\begin{figure}
\vspace{2cm}
    \centering
   \rotatebox{0}{\resizebox{.5\textwidth}{!}{\includegraphics{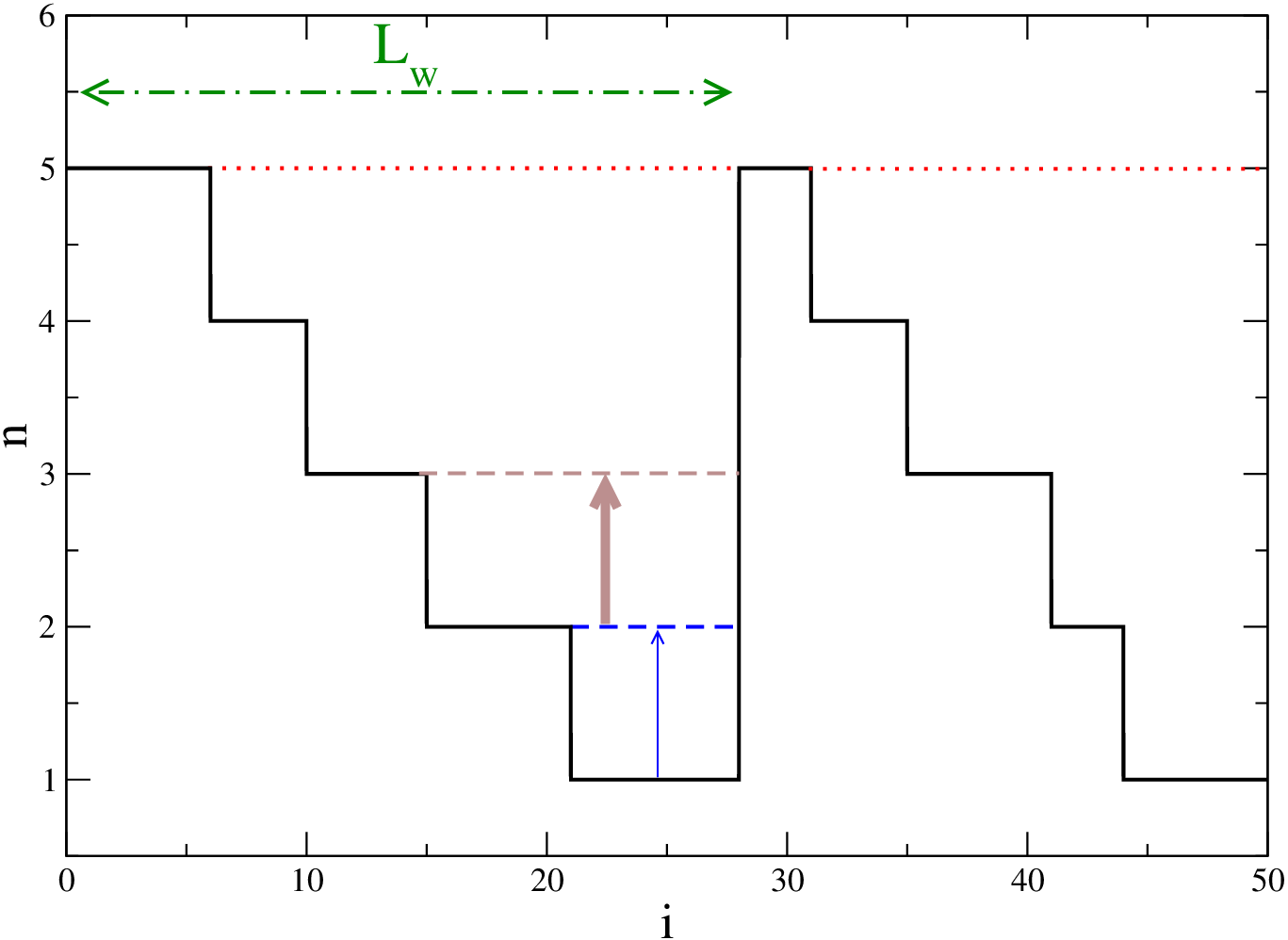}}}
    \caption{(Color online) Schematic representation of a generalized texture 
            for $p=5$.}
\label{figture}
\vspace{2cm}
\end{figure}

For finite values of $2<p<\infty $ we generalize the definition of a texture as
a region of the lattice of length $L_w(t)$ where $p$ subsequent domains are found,
each of average length $L_d(t)\sim L_w(t)/p$, such that moving 
along the lattice the value of $n$ follows the sequence 
$n=p,p-1,...,1$. This is schematically shown in Fig.~\ref{figture}. 

\begin{figure}
\vspace{2cm}
    \centering
   \rotatebox{0}{\resizebox{.5\textwidth}{!}{\includegraphics{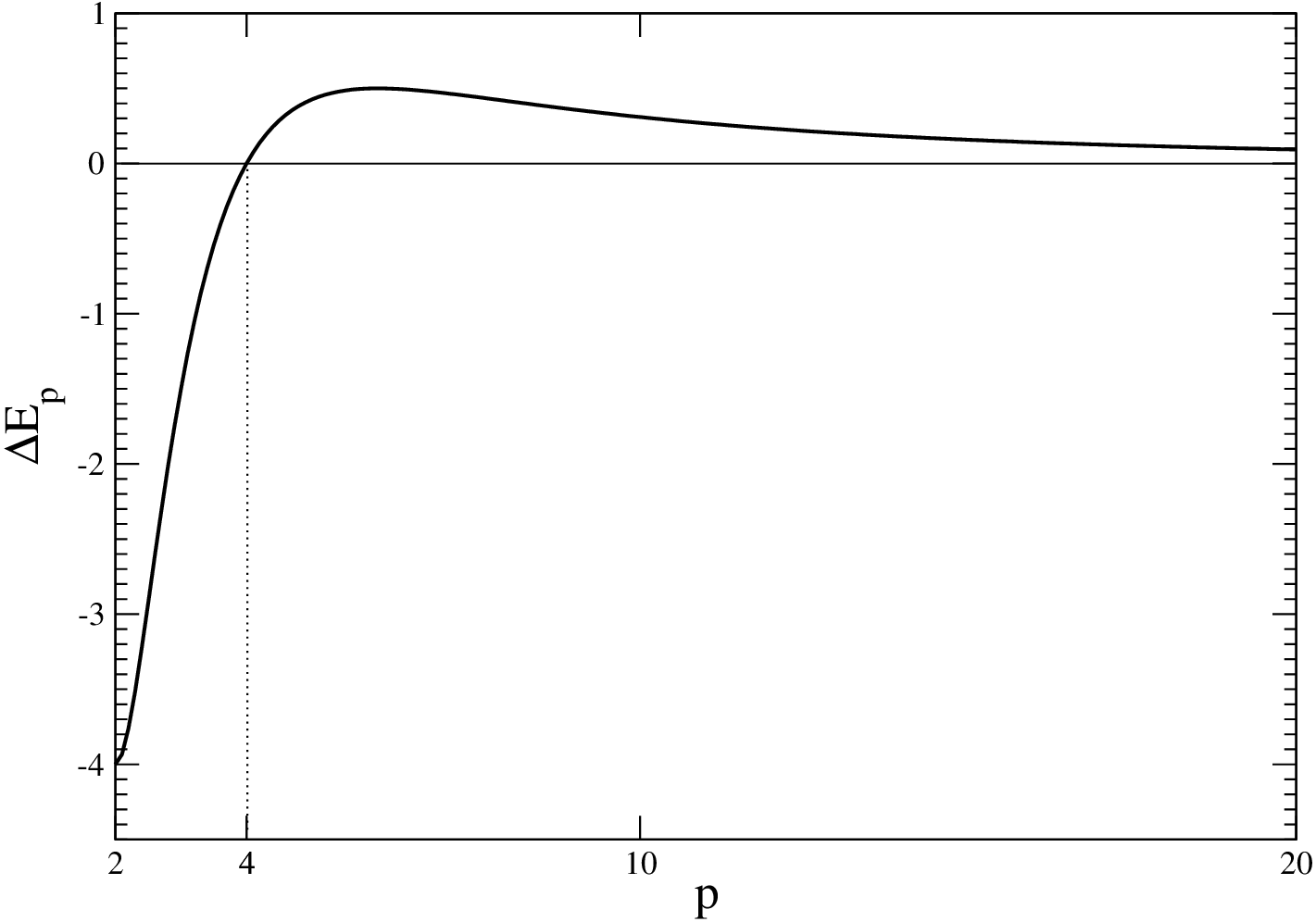}}}
    \caption{The activation energy $\Delta E_p$ needed to destroy textures 
(Eq.~(\ref{activation})) is plotted against $p$.}
\label{figenergy}
\vspace{2cm}
\end{figure}

\begin{figure}
\vspace{2cm}
    \centering
   \rotatebox{0}{\resizebox{.5\textwidth}{!}{\includegraphics{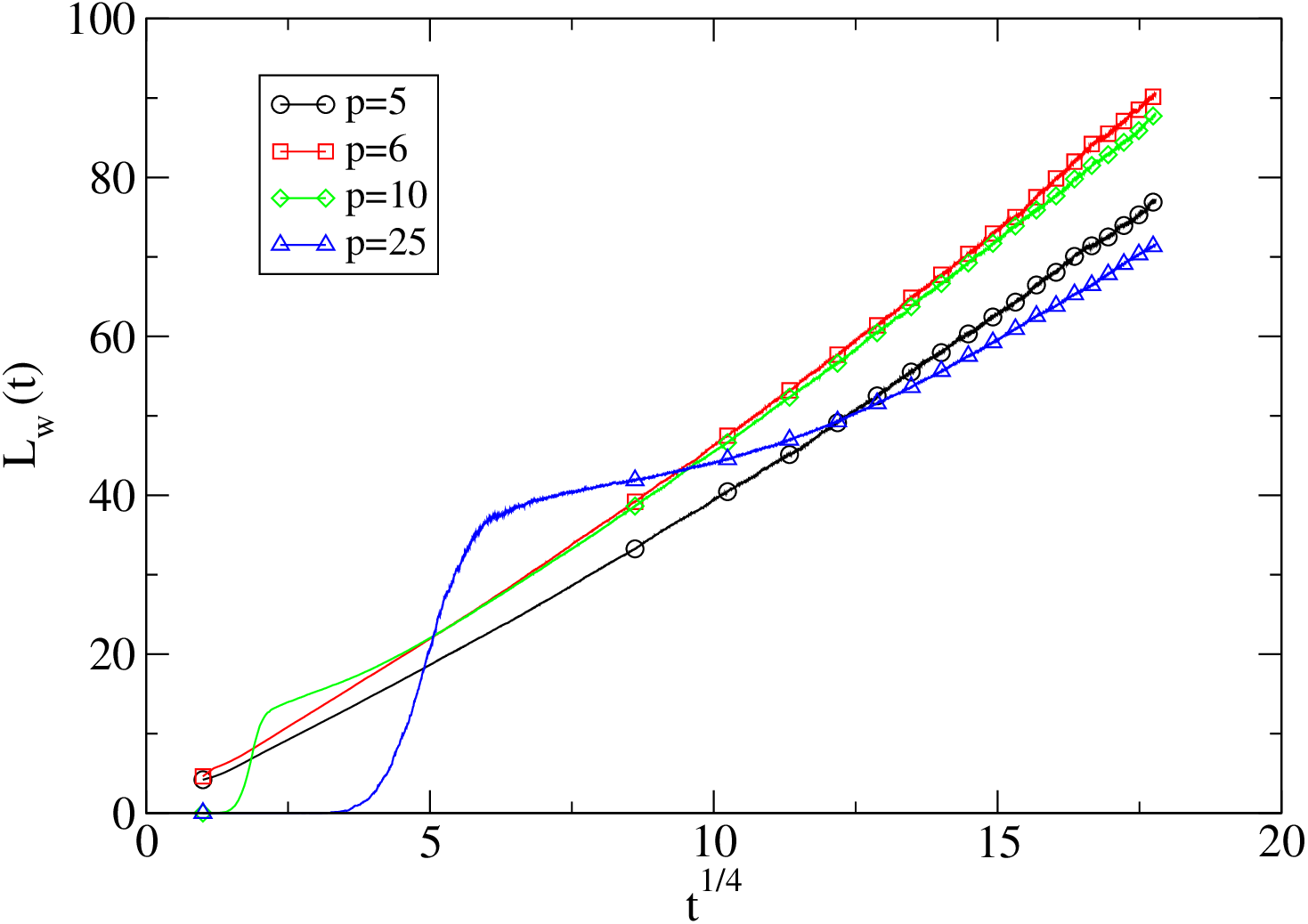}}}
    \caption{(Color online) The average length of textures is plotted against $t^{1/4}$.}
\label{figlengthture}
\vspace{2cm}
\end{figure}

To developed order two mechanisms are possible:
A texture can grow by increasing the number of spins $L_d(t)$ 
on every step. We anticipate that this process 
is found to be relevant for $p>p_c$ and leads
to the power law behavior~(\ref{ldit}) of $L_w(t)$ with $z=4$, as in the XY model.
This behavior competes with the tendency 
to build the largest possible domains, instead of textures.
This amounts to replace a texture with a number $N_D \ll p$ of domains 
each characterized by a single value of $n$.  
However, for $p>p_c$ at $T=0$, 
once textures are present, this process is not allowed. 
In fact, let us consider the situation of Fig.~\ref{figture}
and the possibility to form, in this region, a unique domain with, say, $n=p$
(the dotted line in Fig.~\ref{figture}). 
There are several ways to do this. Suppose one starts by rotating the spins 
with $n=1$
to $n=2$, as shown by the thin arrow in Fig.~\ref{figture}. 
After the move the energy would change by an amount 
\be
\Delta E_p= J[2\cos (2\pi /p) - \cos (4\pi /p)-1].
\label{activation}
\ee
This function is plotted in Fig.~\ref{figenergy}. 
Interestingly  
one has $\Delta E_p\le 0$ or $\Delta E_p>0$ for $p\le p_c$ or $p>p_c$, respectively.
At $T_f=0$ moves with $\Delta E_p>0$ are forbidden. Therefore, for $p>p_c$ 
there is no possibility
to destroy the textures and form domains.
Other possible moves, as, for instance, a rotation from $n=1$ to $n=3$,
correspond to a larger activation energy and are forbidden as well.
Therefore, for $p> p_c$ textures and antitextures are stable against domain formation
and the only ordering mechanism
left is their growth and annihilation, much in the same way as in the
XY model, leading to $z=4$.  
Conversely, for $p\le p_c$ textures
are removed and domains are created whose competition leads to the Ising like
behavior $z=2$. 
As already discussed, in the XY model the exponent $z=4$ is
associated to the growth of the size of single textures. In order to check 
if the same mechanism is at work also in the clock model, in the numerical
simulation we have identified the textures
present in the system at each time and we have computed their average 
size $L_w(t)$.
The results are shown in Fig.~\ref{figlengthture} for different values of $p>p_c$,
showing that, actually,
the size of textures grows as a power law $L_w\sim t^{1/z}$ with $z$
quite compatible with $z=4$ (best fits yield $1/z=0.29\pm 0.02, 0.29\pm 0.02, 
0.28\pm 0.02, 0.23\pm 0.02$ for $p=5,6,10,25$,
respectively).
This confirms that the exponent $z=4$ of the algebraic growth of $L_G(t)$ 
is determined by the texture mechanism, as in the XY model.

\begin{figure}
\vspace{2cm}
    \centering
   \rotatebox{0}{\resizebox{.5\textwidth}{!}{\includegraphics{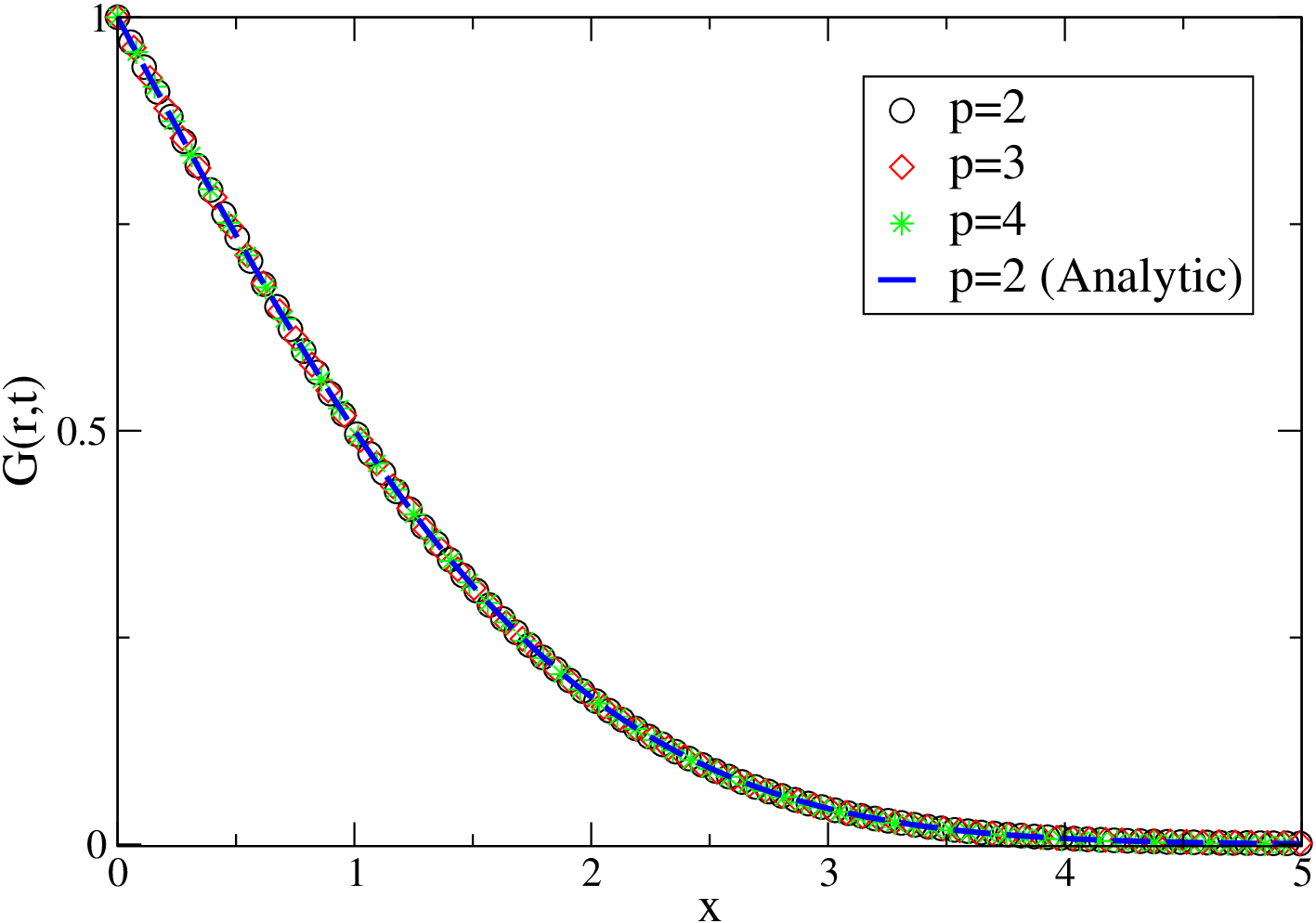}}}
    \caption{(Color online) The correlation function $G(r,t)$ is plotted against $x=r/L_G(t)$ 
             for $p=2,3,4$ at $t=1800$. The dashed line is the analytic 
             expression~(\ref{struttising}).}
\label{figg1}
\vspace{2cm}
\end{figure}

\begin{figure}
\vspace{2cm}
    \centering
   \rotatebox{0}{\resizebox{.5\textwidth}{!}{\includegraphics{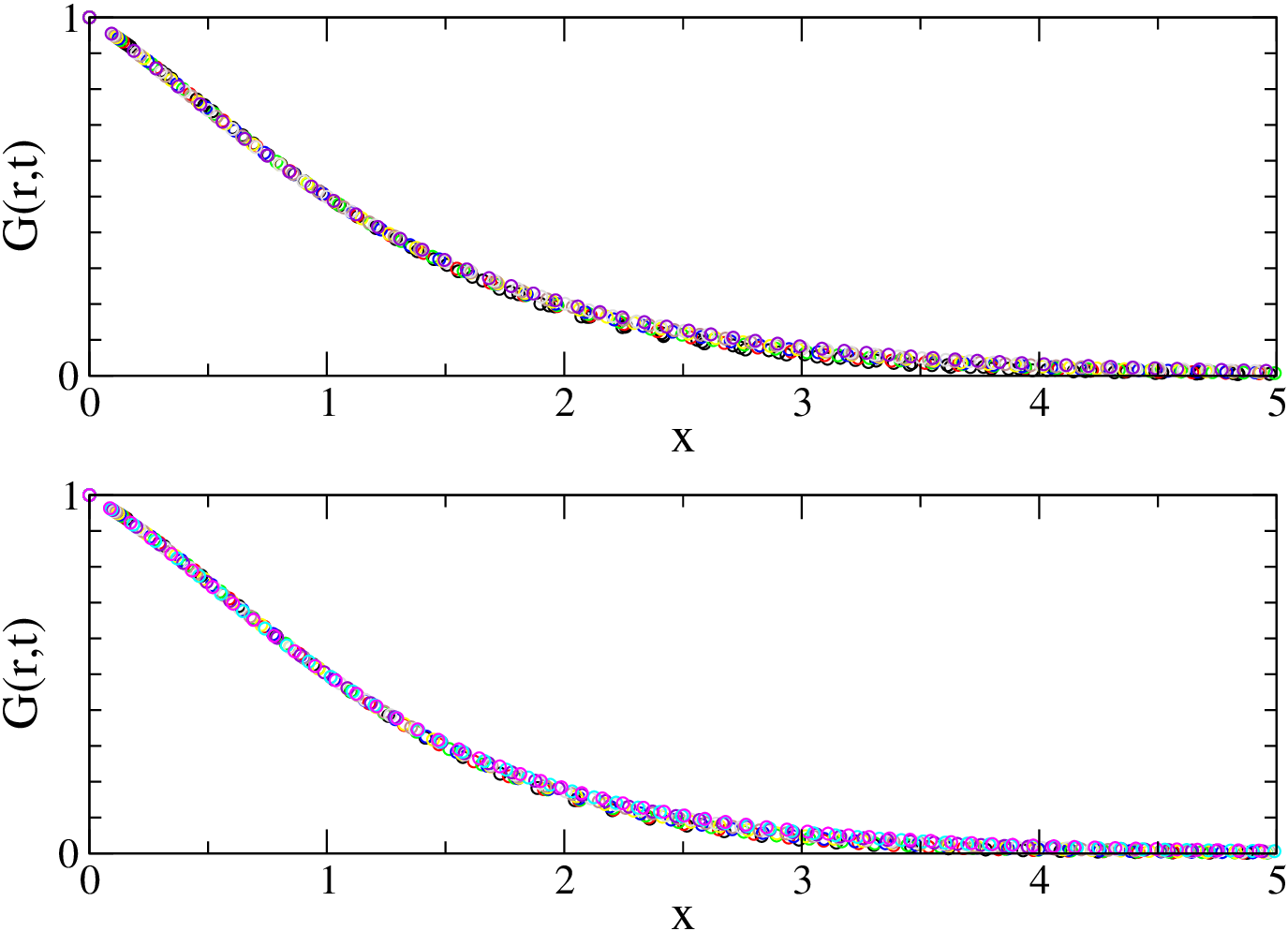}}}
    \caption{(Color online) Data collapse of the correlation function $G(r,t)$
             plotted against $x=r/L_G(t)$ 
             for $p=5$ (upper  panel) and $p=6$ (lower panel), at different
             times ($t=190,245,315,405,520,665,855,1100,1400,1800$).}

\label{figg2}
\vspace{2cm}
\end{figure}

\begin{figure}
\vspace{2cm}
    \centering
   \rotatebox{0}{\resizebox{.5\textwidth}{!}{\includegraphics{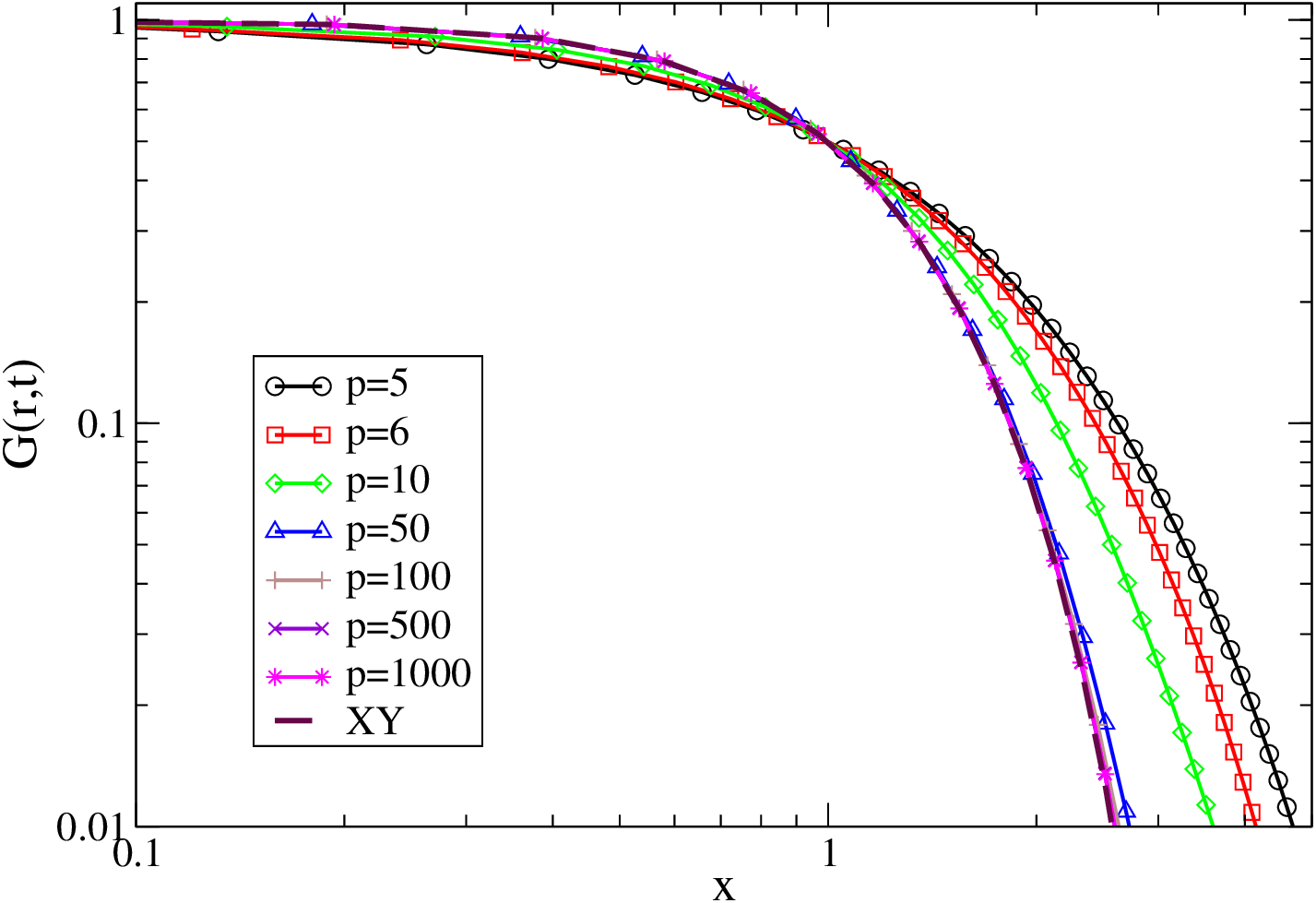}}}
    \caption{(Color online) The correlation function $G(r,t)$ is plotted against $x=r/L_G(t)$ 
             for different values of $p$, at $t=1800$.}

\label{figg3}
\vspace{2cm}
\end{figure}

The previous results for $L_G(t)$ indicate the presence of a crossover at $p=p_c$ from the Ising 
to the XY non-equilibrium universality class. In order to substantiate this
conjecture we have computed other dynamical quantities. 
The equal-time correlation function is
plotted in Figs.~\ref{figg1},\ref{figg2},\ref{figg3} 
against $x=r/L_G(t)$. 
In Fig.~\ref{figg1} the cases with $p=2,3,4$ are considered.
According to Eq.~(\ref{scalgferro}) for $p=2$ one should find collapse 
of the curves with different $s$ on a single mastercurve $g(x)$ 
given by Eq.~(\ref{struttising}). 
This is indeed observed in Fig.~\ref{figg1}.
According to our hypothesis the same behavior
should be observed also for $p=3,4$, as can be verified in the figure.
Moreover, one also finds that the mastercurves $g(x)$
are numerically indistinguishable for different $p$, and they all coincide
with that of Eq.~(\ref{struttising}). This result is trivial
for $p=4$, since in this case the clock model can be mapped
exactly on two non-interacting Ising models. 
The same property could be expected also for $p=3$. 
In fact, by considering $G(r,t)$, 
it easy (see Appendix) to check that 
\be
G(r,t)=\frac{9}{2}G_P(r,t)-\frac{1}{2},
\label{maj1}
\ee
where $G_P(r,t)$ is the {\it single phase } equal time correlation
function of the 3-state Potts model. This quantity was computed in~\cite{Sire},
where it was found
\be
G_P(r,t)=\frac{2}{9}G_I(r,t)+\frac{1}{9},
\label{maj2}
\ee
where $G_I(r,t)$ is the equal time correlation function of
the Ising model.
Plugging Eq.~(\ref{maj2}) into Eq.~(\ref{maj1}) one finds
$G(r,t)=G_I(r,t)$.
The same argument shows also the identity between the two time
correlation functions of the clock model with $p=3$ and the Ising model,
strongly suggesting the complete equivalence between these models.

Let us emphasize that
this result indicates a stronger similarity among the cases $p=2,3,4$ than
a unique non-equilibrium universality class would imply, 
since not only the exponents are equal but the whole functional form of 
the scaling function.   
This results are in contrast with those of ref.~\cite{Liu93} where an
approximate theory was used to show the dependence of $g(x)$ on $p$.
However, the approximation used in~\cite{Liu93} is expected
to improve increasing the dimensionality $d$.

The cases with $p>p_c$ are shown in Fig.~\ref{figg2},\ref{figg3}.
As discussed in Section~\ref{model}, $G(r,t)$ obeys the scaling form~(\ref{scalgferro})
also in the XY model, although dynamical scaling is violated. According to our conjecture,
for $p>p_c$ we expect the same behavior. In Fig.~\ref{figg2} it is shown that,
indeed, the curves at different times collapse when plotted against $x=r/L_G(t)$.
However, differently from the cases $p\le p_c$, the masterfunction $g(x)$ depends
on $p$ and converges to the form~(\ref{xystrutt}) of the XY model for
$p\to \infty $, as shown in Fig.~\ref{figg3}.  

\begin{figure}
\vspace{2cm}
    \centering
   \rotatebox{0}{\resizebox{.5\textwidth}{!}{\includegraphics{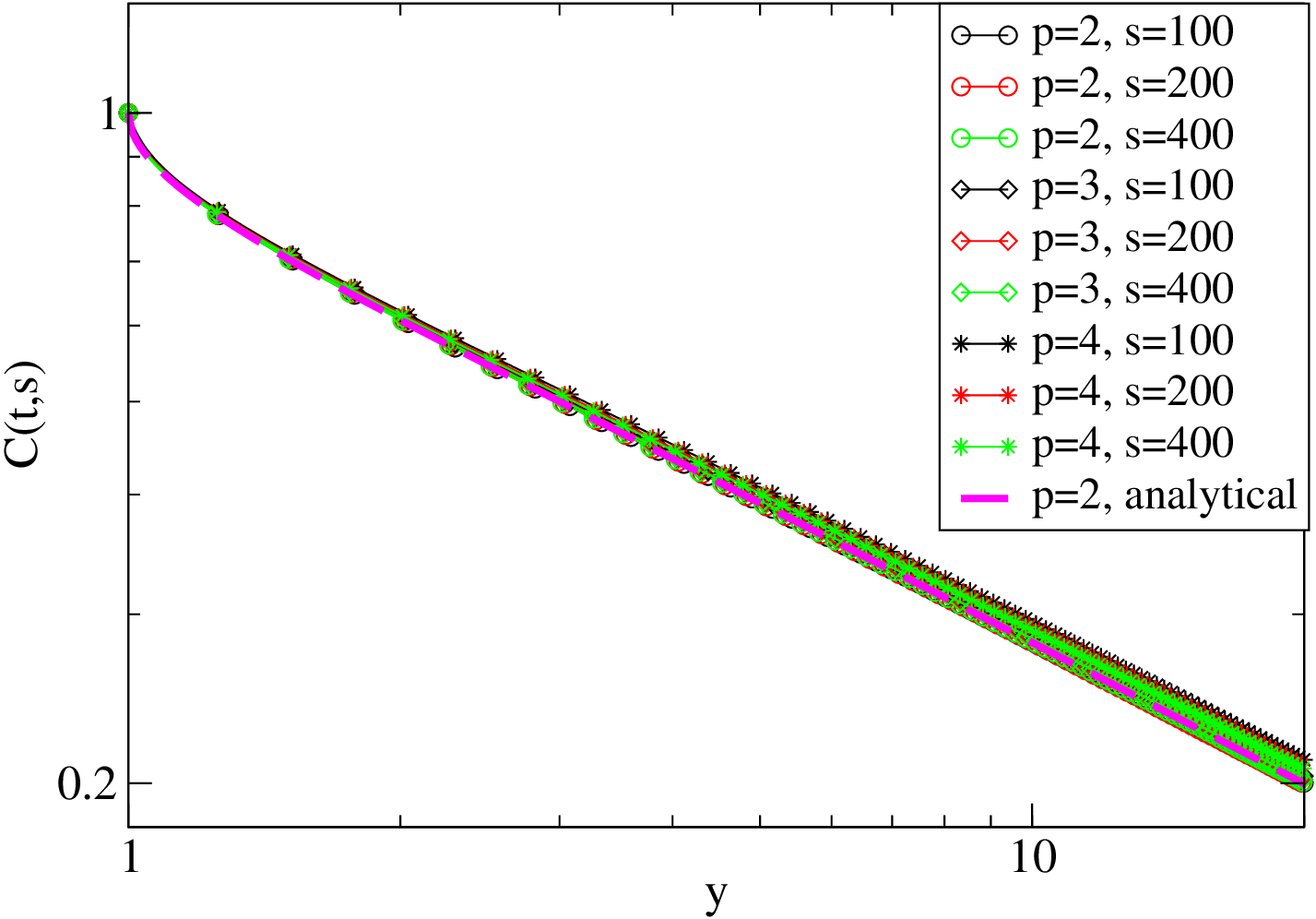}}}
    \caption{(Color online) The autocorrelation function is plotted against $y$ for
             $p=2,3,4$. The dashed line is the analytic expression~(\ref{ccsc}).}
\label{figauto1}
\vspace{2cm}
\end{figure}

\begin{figure}
    \centering
   \rotatebox{0}{\resizebox{.5\textwidth}{!}{\includegraphics{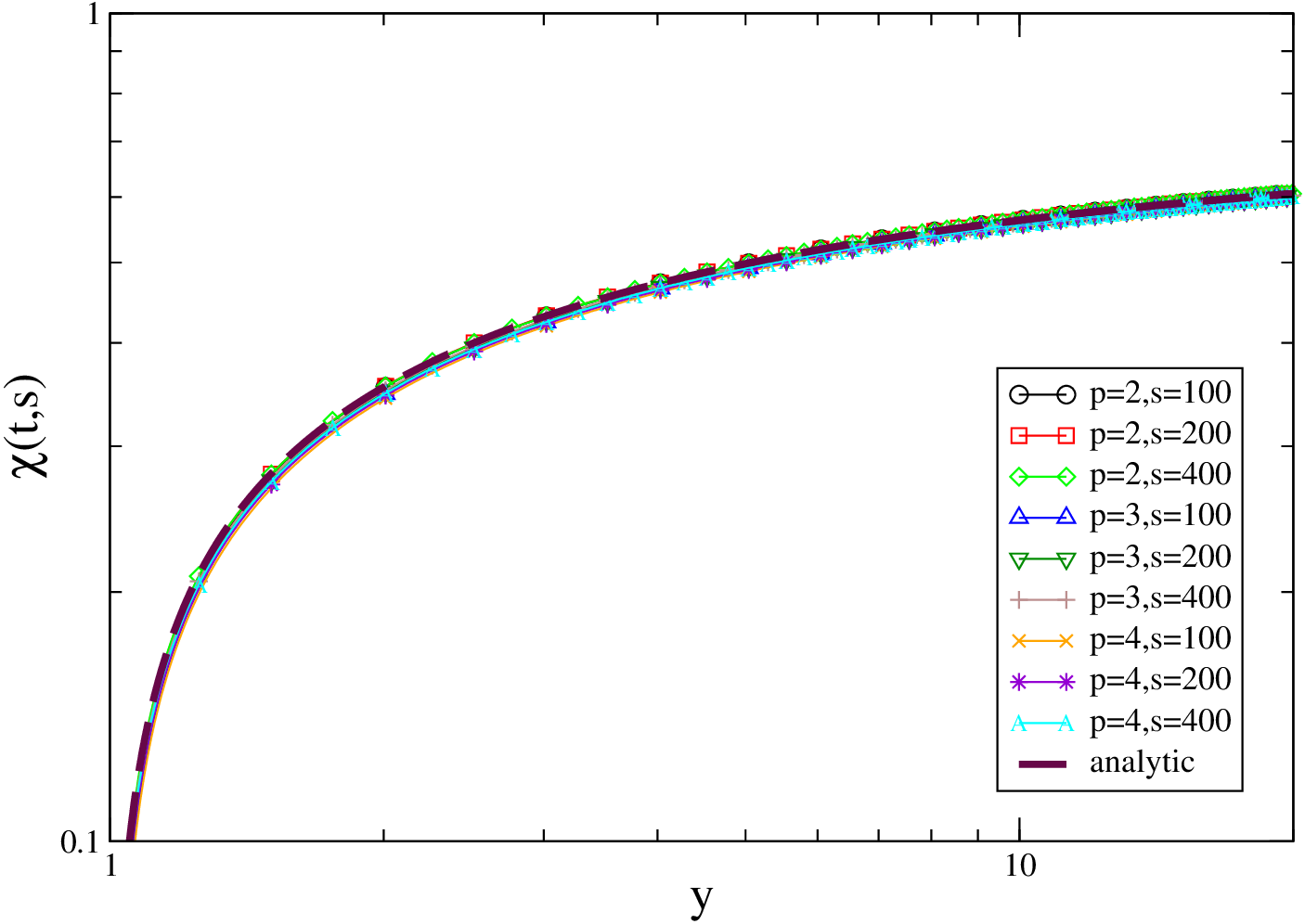}}}
    \caption{(Color online) $\chi (t,s)$ is plotted against $y$ for $p=2,3,4$. 
     The dashed line is the analytic expression~(\ref{ccschi}).}
\label{figchi1}
\vspace{2cm}  
\end{figure}

\begin{figure}
\vspace{2cm}
    \centering
   \rotatebox{0}{\resizebox{.5\textwidth}{!}{\includegraphics{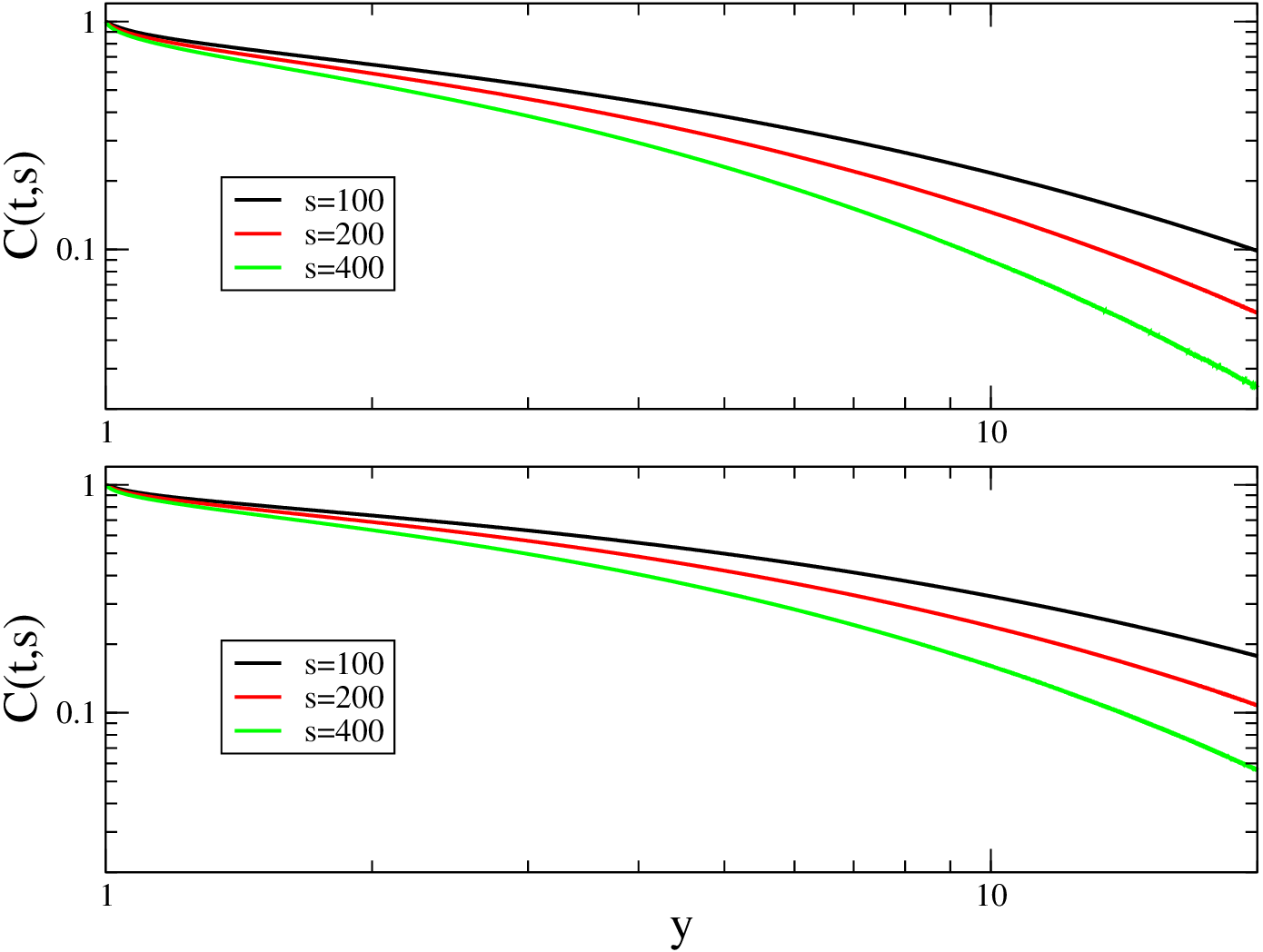}}}
    \caption{(Color online) The autocorrelation function is plotted against $y$ for
             $p=5$ (upper panel) and $p=6$ lower panel).}
\label{figauto2}
\vspace{2cm}
\end{figure}

Let us turn to consider the autocorrelation function, that is
plotted in Figs.~\ref{figauto1}-\ref{figauto2} against $y=t/s$. 
In Fig.~\ref{figauto1} the cases with $p=2,3,4$ are considered.
Here the situation is analogous to that of $G(r,t)$.
For $p=2$ one should find collapse 
of the curves with different $s$ on a mastercurve $h(y)$, Eq.~(\ref{ccs}). 
This is indeed observed in Fig.~\ref{figauto1}.
The same behavior
is observed also for $p=3,4$.
Again, as for $G(r,t)$, we find that the mastercurves $h(y)$
are numerically indistinguishable for different $p$, and they all coincide
with that of Eq.~(\ref{ccsc}). 
In order to check if this property is completely general, namely if 
every observable is characterized by the same exponents and scaling
functions for $p=2,3,4$, besides the correlation functions we have also computed the 
integrated autoresponse function  
\be
\chi (t,s)=\int _s ^t dt' R(t,t').
\label{integrated}
\ee
Here
\be
R(t,t')=\sum _{\alpha} \left . \frac {\partial \langle \sigma_i^\alpha (t) \rangle}
{\partial h_i^\alpha (t')}\right \vert _{\vec h_i=0},
\ee
$\alpha =1,2$ being the generic vector components, 
is the linear autoresponse function associated 
to the perturbation caused by an impulsive magnetic field $\vec h _i$
switched on at time $t'<t$.
In the Ising model~\cite{Lippiello00}, in the $T\to 0$ limit
one finds
\be
\chi (t,s)=f(t/s),
\label{ccschichi}
\ee
with
\be
f(y)=\frac{1}{\sqrt 2} \left (1-\frac{2}{\pi}\arcsin \sqrt {y^{-1}}\right ).
\label{ccschi}
\ee
Here we measure the
response function using the efficient method derived in~\cite{Lippiello05}
without applying the perturbation.

The behavior of $\chi (t,s)$ is shown in Fig.~\ref{figchi1} 
for the cases $p=2,3,4$.
One finds collapse
of the curves with different $s$ on a mastercurve $f(y)$,
as in Eq.~(\ref{ccschichi}) for $p=2$. 
Also in this case mastercurves $f(y)$ for different $p$ are
numerically indistinguishable. In conclusion, then, our data for
$G(r,t)$, $C(t,s)$ and $\chi (t,s)$ confirm that 
the cases with $p=2,3,4$ share the same exponents and scaling functions.
Notice that having the same scaling function both for $C(t,s)$ and
$\chi (t,s)$, the cases with $p\le p_c$ have also the same parametric 
plot of $\chi (t,s)$ versus $C(t,s)$~\cite{Lippiello00}.
 
\begin{figure}
    \centering
   \rotatebox{0}{\resizebox{.5\textwidth}{!}{\includegraphics{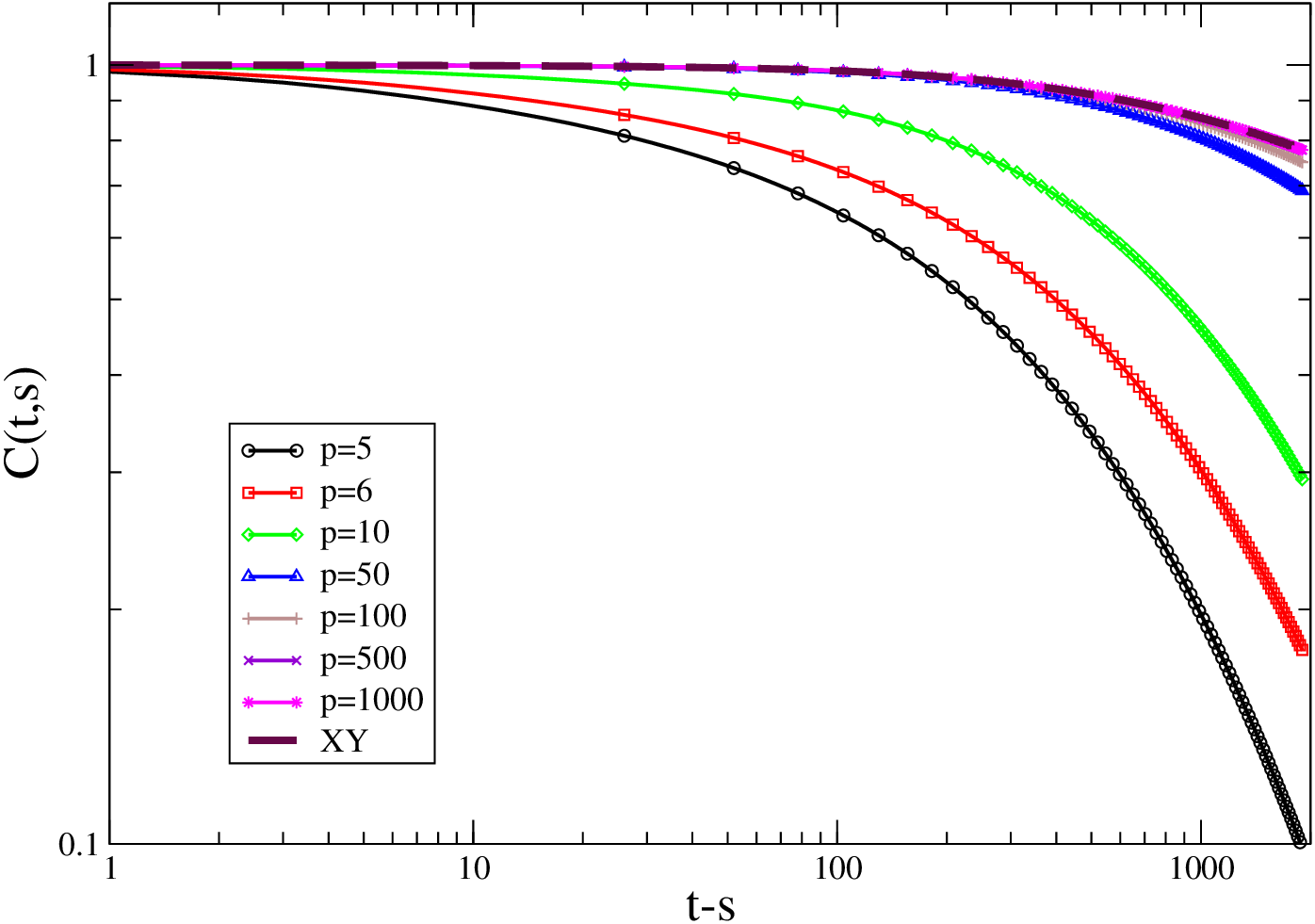}}}
    \caption{(Color online) The autocorrelation function is plotted against $y$ for $s=100$
             and several values of $p$.}
\label{figauto3}
\vspace{2cm} 
\end{figure}

The situation is radically different for $p>p_c$.
We expect here to see a texture-dominated XY-like dynamics,
with violations of dynamical scaling that can be detected from
$C(t,s)$.
In fact, this is what one observes in
Fig.~\ref{figauto2}, where the autocorrelation function is plotted against $y$. 
For each value of $p$, curves with different
values of $s$ do not collapse.
The whole behavior is qualitatively similar to that of the XY model described
by Eq.~(\ref{cxy}), which predicts the lowering of the curves for
fixed $y$ as $s$ increases. 
Quantitatively, as already observed regarding $G(r,t)$, the analytic
form of the curves depends on $p$ and is different from that
of the XY model, namely Eq.~(\ref{cxy}). 
As shown in Fig.~\ref{figauto3}, Eq.~(\ref{cxy}) is gradually approached
increasing $p$.

\subsection{Quenches to $T_f>0$.} \label{Tnzer}

When quenches to finite temperatures are considered, as already discussed
in Sec.~\ref{model}, one has a finite equilibration time $\tau _p^{eq}(T_f)$.
In the following we will always discuss the ordering kinetics preceding
the equilibration time, namely for $t\ll \tau _p^{eq}(T_f)$.

According to our hypothesis, the XY-like behavior observed for $p>p_c$ is due to
the impossibility to eliminate textures and form domains, because this would 
require activated processes with $\Delta E_p>0$ given by Eq.~(\ref{activation}).
Quenching to a finite temperature those processes are no longer forbidden 
and we expect textures to start being removed after a characteristic
time $\tau _p ^{cross}(T_f)$. In order to estimate the crossover time let us
consider again the situation of Fig.~\ref{figenergy}. The activated process 
described by the thin arrow, where the spins with $n=1$ are rotated to $n=2$, 
is a first action towards the removal of the texture, but the
texture is not disappeared yet. The second action is the
rotation of spins from $n=2$ to $n=3$, indicated by a bold arrow in the figure~\cite{nota1}.
This requires an energy
\be
\Delta E_p^{(2)}= J[\cos (2\pi /p) +\cos (4\pi /p) - \cos (6\pi /p)-1].
\label{activation2}
\ee
Then a third action is required, where spins with $n=3$ are rotated to $n=4$, and 
so on, until, after $p-1$ steps all the spins in the region considered have $n=p$.
It is easy to generalize Eqs.~(\ref{activation},\ref{activation2}) to
the generic $m$-th action:
\be
\Delta E_p^{(m)}= J[\cos (2\pi /p) +\cos (2m\pi /p) - \cos (2(m+1)\pi /p)-1].
\label{activationm}
\ee
Let us consider $\Delta E_p^{(2)}$. This quantity is positive for $p>6$. 
For $p=5,6$, therefore, the second
action is not an activated process, while it is activated for $p>6$.
In general, from Eq.~(\ref{activationm}) one has $\Delta E_p^{(m)}>0$ for $p>2+2m$. 
The accomplishment of an action requires a time~\cite{nota2}
\be
t _p ^{(m)}(T_f)\simeq \left [ w_p(\Delta E_p/T_f) \right ]^{-1}=
\frac{2}{p}\left \{1+\exp [\Delta E_p^{(m)}/T_f]\right \},
\label{ttcross}
\ee
$w_p$ being the transition rates defined in Eq.~(\ref{metropolis}).
The crossover time, namely the characteristic time after which textures are removed,
is given by the sum of the times required for all the $p-1$ actions. It can be evaluated as
\be
\tau _p ^{cross}(T_f)=\sum _{m=1}^{p-1}t _p^{(m)}(T_f).
\label{tcross}
\ee
In the limit $T_f\to 0$ the sum is dominated by the process with the
largest activation energy 
\be
\tau _p ^{cross}(T_f\simeq 0)=Sup _{\{m=1,p-1\}}t _p^{(m)}(T_f).
\ee
The $Sup $ in this equation is obtained for $m=m^*$ given by  
\be
m^* = \left \{ \begin{array}{ll}
        1  \qquad $for$ \qquad p < 10   \\
	\left [\frac {p-2}{4} \right ]   \qquad $for$ \qquad  p\ge 10 
        \end{array}
        \right .
\ee
where $[ x ]$ is the integer part of $x$.
Then, in the low-$T$ limit one has
\be
\tau _p ^{cross}(T_f\simeq 0)=t _p^{m^*}(T_f).
\ee
In conclusion,
for $p\le p_c$ no activated processes are required and the system
immediately enters the Ising-like phase ordering behavior. 
For $p>p_c$, instead, the dynamics is initially of the XY type until,
at $t\sim \tau _p^{cross}(T_f)$ there is
a crossover to the Ising-like non-equilibrium behavior. 

\begin{figure}
    \centering
   \rotatebox{0}{\resizebox{.5\textwidth}{!}{\includegraphics{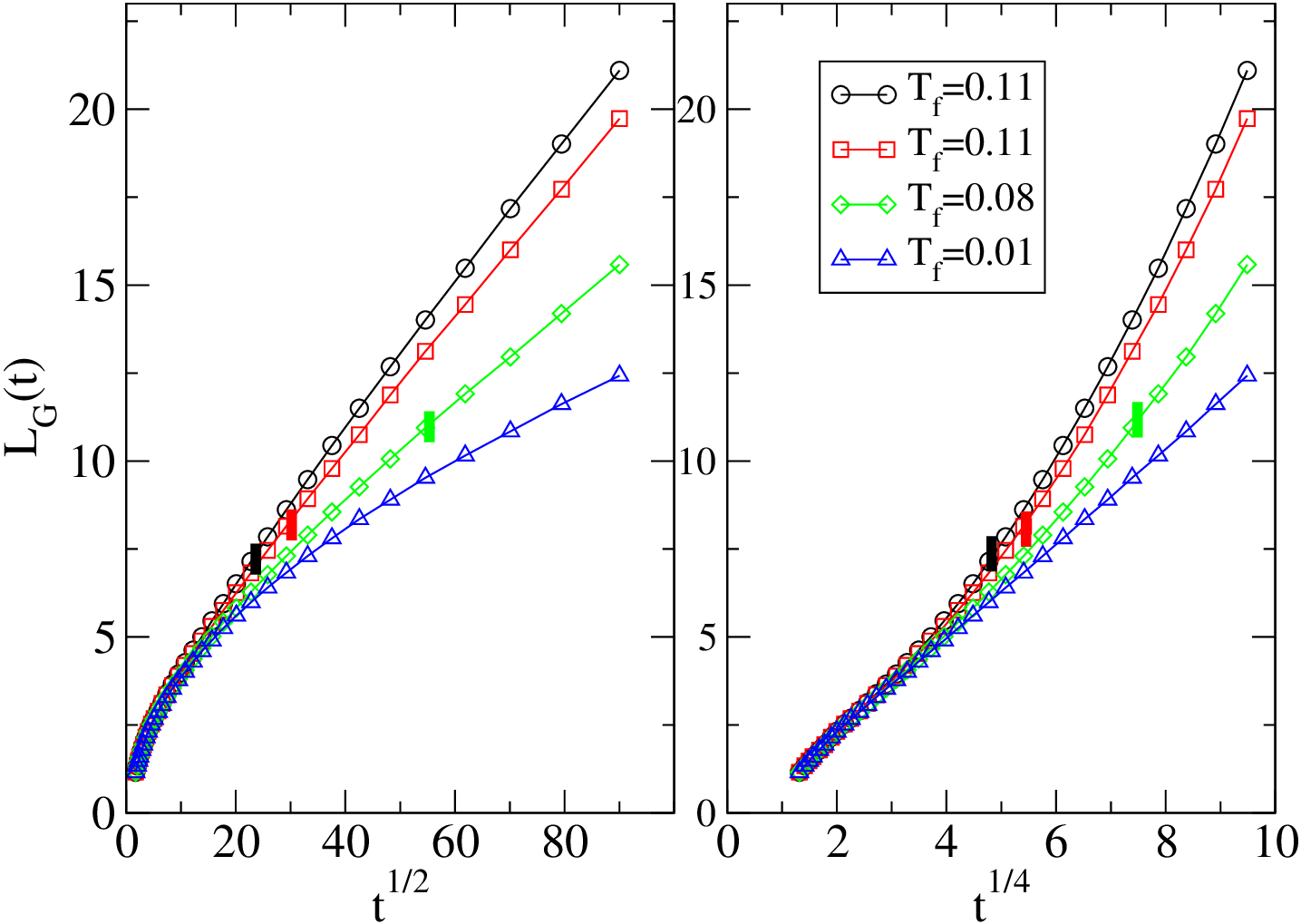}}}
    \caption{(Color online) $L_G(t)$ is plotted against $t^{1/4}$ (left panel) or
            versus $t^{1/2}$ (right panel), for a quench of a system with
            $p=6$ and different values of $T_f$.
            Vertical segments on the curves for different $T_f$ represent 
            $\tau _p^{cross} (T_f)$ obtained from Eq.~(\ref{tcross}) as discussed
            in the text.
            For the smallest temperature $\tau _p^{cross} (T_f)$ is outside the range of times
            of the figure.}
\label{figlengthT2}
\vspace{2cm}
\end{figure}

\begin{figure}
    \centering
   \rotatebox{0}{\resizebox{.5\textwidth}{!}{\includegraphics{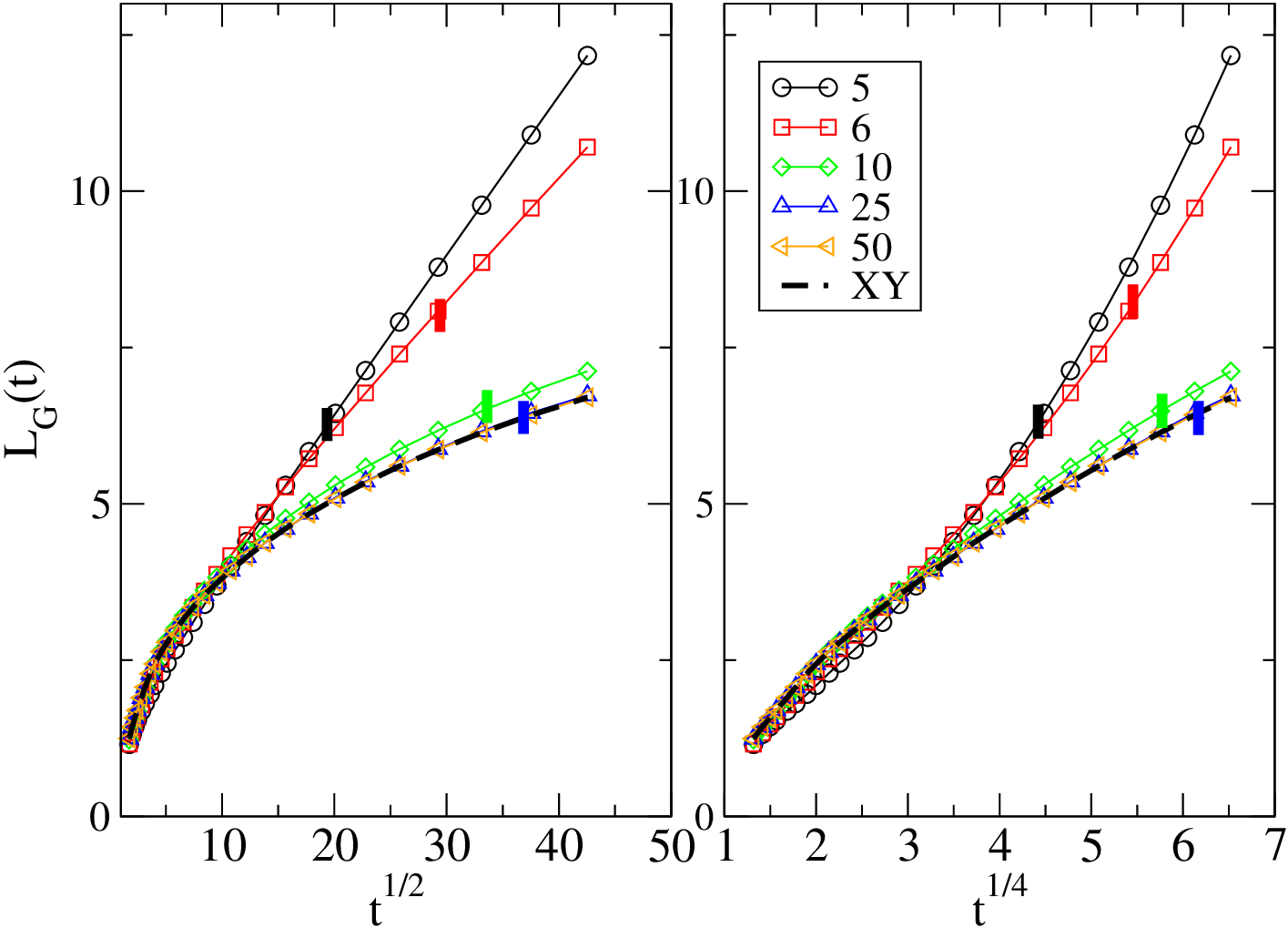}}}
    \caption{(Color online) $L_G(t)$ is plotted against $t^{1/4}$ (left panel) or
            versus $t^{1/2}$ (right panel), for a quench at $T_f=0.1$ and several
            values of $p>p_c$.
            Vertical segments on the curves for different $p$ represent $\tau _p^{cross} (T_f)$.
            For $p>25$ $\tau _p^{cross} (T_f)$ is outside the range of times
            of the figure.}
\label{figlengthT1}
\vspace{2cm}
\end{figure}

The crossover 
can be appreciated in Figs.~\ref{figlengthT2},\ref{figlengthT1}.
The former shows the behavior of $L_G(t)$ for $p=6$
and different values of $T_f$.
Here one observes initially the same behavior as for $T_f=0$,
namely $L_G(t)\propto t^{1/4}$, i.e. a straight line in the plot
of $L_G(t)$ against $t^{1/4}$ (right panel). For larger times
there is a crossover to the Ising behavior $L_G(t)\simeq t^{1/2}$,
namely a straight line in the plot of $L_G(t)$ versus $t^{1/2}$ (left panel).
Although the crossover is a quite smooth phenomenon, as can be
seen in Fig.~\ref{figlengthT2}, $\tau _p^{cross}(T_f)$ given by Eq.~(\ref{tcross}),
represented by thick segments across the lines,
turns out to be of the correct order of magnitude for all the temperatures
considered.

In Fig.~\ref{figlengthT1} we plot $L_G(t)$ for $T_f=0.1$ and different values of $p$.
One observes the same pattern of behavior of Fig.~\ref{figlengthT2} with a
crossover from a power law growth with $z=4$ to one with $z=2$. 
The crossover time~(\ref{tcross}) grows with $p$, as expected.
 
\section{Conclusions} \label{concl}

In this paper we have studied the phase-ordering kinetics of the one dimensional 
$p$-state clock model. We have shown the existence of a critical value $p_c=4$
separating two radically different dynamical behaviors. For $p\le p_c$
the dynamics is in all respects analogous to that of the Ising model
with $p=2$. Phase-ordering proceeds by means of formation and subsequent
growth of domains through interface diffusion and annihilation.
This similarity goes beyond the qualitative level: we find the same
exponent and scaling functions for every $p\le p_c$ and for all the 
one-time or two-time quantities considered. This reflects
a deeper similarity than what a unique universality class, involving
only the value of the exponents, would imply. 
For $p>p_c$ the dynamics changes dramatically, due to the relevant
role played by textures. While for $p\le p_c$ textures are quickly removed
by means of non-activated processes, for $p>p_c$ their
removal can only be realized through activated processes.
For quenches to $T_f=0$, activated process are forbidden, and, therefore,
textures remain in the system up to the longest times. Their
peculiar growth mechanisms characterize the dynamics, similarly
to what happens in the one-dimensional XY model, with the notable feature of
violation of dynamical scaling and the anomalous growth with $z=4$ of the
winding length $L_w(t)$. For quenches to finite $T_f$,
textures survives up to a characteristic time $\tau _p^{cross}(T_f)$ which
can be rather long for small temperatures or large $p$. A crossover phenomenon
is then observed from an initial dynamics of the XY type, to
a later Ising-like behavior.

Our results are at odd with what is found in Ref.~\cite{Liu93} 
where an approximate analytical
solution of the clock model in arbitrary dimension is obtained, finding 
an analogous scaling behavior for all $p< \infty $ but with 
$p$-dependent scaling functions. In the present one-dimensional case,
instead, the situation is the opposite. There is not an analogous scaling 
behavior for all values of $p$, but a qualitative difference occurs 
crossing $p_c$. In addition, when scaling holds, namely for $p\le p_c$,
the scaling functions do not depend on $p$.
We believe, however, the behavior of the system considered
in this paper, to be peculiar. Actually, the different dynamics
observed crossing $p_c$ is determined by the simultaneous presence of
interfaces and textures. On the basis of the discussion of Sec.~\ref{intro}
we expect a similar situation to be only
realized in $N$-component vectorial models with discrete states
and $N=d+1$, where extended defects without a core may exist.
For instance, it would be very interesting to study if a similar pattern
is observed in $d=2$ for a generalization
of the clock model where a three component order parameter is only allowed to
point on a finite number $p$ of directions. 
In addition, we expect the remarkable feature of unique scaling
functions for different values of $p$ to be
peculiar to the one-dimensional case. Considering the function
$G(r,t)$, for instance, the scaling function describes the spatial distribution
of domains and it is quite evident that in $d>1$ this depends on
$p$. Taking the case $d=2$, for simplicity, one has the
usual bicontinuous domain structure of domains and interfaces for $p=2$, while
for $p>2$ there is a different pattern with interfaces and 
vortices~\cite{Kaski83}. However, in the one dimensional case
interfaces are point-like objects for all values of $p$ 
and one does not expect relevant differences in their spatial distribution 
when $p$ is changed. 

Finally, it would be very interesting to study if a similar pattern is observed
in the one-dimensional clock model with a conserved order parameter.
Concerning the value of the growth exponent $z$, which in the non-conserved case 
considered here effectively  
discriminate the Ising dynamics with $z=2$ from the XY behavior with $z=4$,
in the conserved case one should observe a 
crossover from $z=3$ to $z=6$~\cite{Bray94,Rutenberg95}. 

{\bf Acknowledgment}

We acknowledge the referee for valuable suggestions.
\vspace{.6cm}

This work has been partially supported
from INFM through PAIS and from MURST through PRIN-2004.

\vspace{1cm}
{\noindent \bf APPENDIX }
\vspace{.6cm}
\appendix \label{appendix1}

For the 3-states clock model the correlation between two spins 
at a certain time $t$ can be written as
\be
G(r,t)=\langle \sigma _i \sigma _j\rangle = 
\sum _{n,n'=1,3}\cos [\theta _i (n)-\theta _j(n')]
P_i(n,t)P_{i,j}(n,t\mid n',t),
\ee
where $r$ is the distance between $i$ and $j$. 
$n$, $\theta _i$ (and their relation) are defined in Eq.~(\ref{theta}),
$P_i(n,t)$ is the probability to find the spin on site $i$
in the state $n$ at time $t$, and $P_{i,j}(n,t\mid n',t)$ is the conditional
probability to find the state $n'$ on site $j$ provided that the
state $n$ is found in $i$. 
Isolating the diagonal terms one has
\be
G(r,t) = 
\sum _{n=1,3}P_i(n,t)P_{i,j}(n,t\mid n,t)-\frac{1}{2}
\sum _{n=1,3}P_i(n,t)\sum _{n'\ne n}P_{i,j}(n,t \mid n',t),
\ee
where we have used the value $\cos (\theta _i-\theta _j)=-1/2$ when $\theta _i\ne \theta _j$.
Since $\sum _{n'\ne n}P_{i,j}(n,t\mid n',t)=1-P_{i,j}(n,t\mid n,t)$
one has
\be
G(r,t)= 
-\frac{1}{2}\sum _{n=1,3}P_i(n,t)+
\frac{3}{2}\sum _{n=1,3}P_i(n,t)P_{i,j}(n,t\mid n,t)
=-\frac{1}{2}+
\frac{3}{2}\sum _{n=1,3}P_i(n,t)P_{i,j}(n,t\mid n,t)
\label{appe3}
\ee

Let us turn now to the Potts model where a generic spin on site $i$ can be
found in the states labeled with $m_i=1,2,3$.
Following Ref.~\cite{Sire}, we define an auxiliary
field $\phi _i (n)$ such that $\phi _i (n)=1$ if $m_i=n$,
where $n$ is a reference state,
and $\phi _i (n)=0$ otherwise.
The correlation of the auxiliary field is the  {\it single phase}
correlation function of the  Potts model  and 
can be written as
\be
G_n(r,t)=\langle \phi _i (n)\phi _j (n)\rangle=
P_i(n,t)P_{i,j}(n,t\mid n,t),
\label{appe4}
\ee
where the probabilities are defined analogously to the those of the clock
model introduced above. Recognizing $G_n(r,t)$ in the last term of the
right hand side of Eq.~(\ref{appe3}) one arrives at
\be
G(r,t)= -\frac{1}{2}+
\frac{3}{2}\sum _{n=1,3}G_n(r,t).
\label{resul}
\ee
Because of the  rotational symmetry one has  $G_P(r,t)=G_n(r,t)$ for all values of $n$ and then one recovers Eq.~(\ref{maj1}).


\begin{thebibliography}{99}
\bibitem{nota1}
As for the first action, other possible moves require a larger activation energy
and are, therefore, suppressed at low temperatures.
\bibitem{nota2}
In a texture there may be several adjacent spins with the same value of $n$
on a step, as in Fig.~\ref{figture}. Since the accomplishment of an action requires 
all these spins to be rotated, a number of elementary moves
with $\Delta E=0$ may occur besides the (possibly) activated processes
with energy variation $\Delta E_p ^{(m)}$. 
These moves correspond to the Brownian displacement of 
the boundaries between, say, spins with $n=2$ and $n=3$ in Fig.~\ref{figture}.
They can be disregarded in the computation of $t_p^m(T_f)$ at low
temperatures since they require a microscopic time to occur.
\end{thebibliography}
\end{document}